\documentclass[11pt,reqno]{amsart}
\usepackage{cases}
\usepackage{amsfonts}
\usepackage{txfonts}
\usepackage{stmaryrd}
\usepackage{}
\usepackage{amsmath}
\usepackage{amssymb}

\newtheorem{thm}{Theorem}[section]

\numberwithin{equation}{section}

\newcommand{\mysection}[1]{\section{#1}\setcounter{equation}{0}}

\newfont{\bb}{msbm10 at 12pt}

\def\R{\hbox{\bb R}}

\def\g{{\bar g}}
\def\h{{\bar h}}
\def\k{{\lambda}}

\def\hconn{\breve{\nabla}}
\def\cconn{\hat{\nabla}}

\def\P{\mathcal P}

\def\H{\mathcal H}
\def\cR{\mathcal R}
\def\Lie{\mathcal L}

\newcommand{\bal}{\begin{aligned}}      \newcommand{\eal}{\end{aligned}}
\newcommand{\ba}{\begin{array}}      \newcommand{\ea}{\end{array}}
\newcommand{\bc}{\begin{center}}     \newcommand{\ec}{\end{center}}
\newcommand{\be}{\begin{enumerate}}  \newcommand{\ee}{\end{enumerate}}
\newcommand{\beq}{\begin{eqnarray}}  \newcommand{\eeq}{\end{eqnarray}}
\newcommand{\beQ}{\begin{eqnarray*}} \newcommand{\eeQ}{\end{eqnarray*}}
\newcommand{\bi}{\begin{itemize}}    \newcommand{\ei}{\end{itemize}}
\newcommand{\bt}{\begin{tabular}}    \newcommand{\et}{\end{tabular}}
\newcommand{\bdm}{\begin{displaymath}} \newcommand{\edm}{\end{displaymath}}

\let\pa=\partial

\newcommand{\ls}{\setlength{\baselineskip}{12pt}
                 \setlength{\parskip}{3mm}}

\begin{document}

\title[Negative energy]{spacelike hypersurfaces with negative total energy in de Sitter spacetime}

\author{Zhuobin Liang and Xiao Zhang}
\address{Institute of Mathematics, Academy of Mathematics and
Systems Science, Chinese Academy of Sciences, Beijing 100190, PR
China}
\email{liangzhuobin@amss.ac.cn, xzhang@amss.ac.cn}

\date{}

\begin{abstract}

De Sitter spacetime can be separated into two parts along two kinds of hypersurfaces and the half-de Sitter spacetimes
are covered by the planar and hyperbolic coordinates respectively.
Two positive energy theorems were proved previously for certain $\P$-asymptotically de Sitter and $\H$-asymptotically
de Sitter initial data sets by the second author and collaborators.
These initial data sets are asymptotic to time slices of the two kinds
of half-de Sitter spacetimes respectively, and their mean curvatures are bounded from above by certain constants. While
the mean curvatures violate these conditions, the spacelike hypersurfaces with negative total energy in the two kinds of
half-de Sitter spacetimes are constructed in this short paper.

\end{abstract}

\maketitle \pagenumbering{arabic}


\mysection{Introduction}\ls

In general relativity, the positive energy theorem plays a fundamental role which serves as a consistent verification
of the theory. If the positive energy theorem holds, the spacetime with vanishing total energy can be viewed as the
ground state. In the case of zero cosmological constant, the positive energy theorem for asymptotically
flat spacetimes was firstly proved by Schoen and Yau \cite{SY1, SY2, SY3}, and then by Witten \cite{W} using a different
method. Recently, Witten's method was extended successfully to asymptotically Anti-de Sitter spacetimes and the positive
energy theorem was proved completely and rigorously in the case of negative cosmological constant \cite{Wa, CH, Z3, M, XZ}.

Recent cosmological observations indicated that our universe has a positive cosmological constant. It is therefore important
to study whether the positive energy theorem for asymptotically de Sitter spacetimes holds. De Sitter spacetime with cosmological
constant $\Lambda =\frac{3}{\k ^2}>0$, $(\k>0)$ is a hypersurface embedded into 5-dimensional Minkowski spacetime $\R ^{1,4}$
 \beq
-\left(X ^0\right) ^2 + \left(X ^1\right) ^2+\left(X ^2\right) ^2+\left(X ^3\right) ^2+\left(X ^4\right) ^2 =\frac{3}{\Lambda} \label{dS}
 \eeq
with the induced metric (cf. \cite{HE}). It is covered by global coordinates where each
time slice is a 3-sphere with constant curvature and has no spatial infinity. As the corresponding
Killing vector fields to the Lorentzian generators are timelike in some region of de Sitter
spacetime and spacelike in some other region, there is no positive conserved energy in de Sitter
spacetime \cite{W2, AD}. So the issue of the positive energy theorem for
asymptotically de Sitter spacetimes becomes sophisticated. However, de Sitter spacetime can be separated into
two parts along the hypersurface $X^0=X^4$ and the half-de Sitter spacetime is covered by the planar coordinates
with the de Sitter metric
 \beq \label{dS-planar}
\tilde g ^\P _{dS} = -dt ^2 + e ^\frac{2t}{\k} \left(\left(dx ^1\right)^2 +\left(dx ^2\right)^2
+\left(dx ^3\right)^2\right).
 \eeq
Similarly, de Sitter spacetime can be separated into two parts along the hypersurface $X^4=-\lambda$
and the half-de Sitter spacetime is covered by the hyperbolic coordinates with the de Sitter metric
 \beq \label{dS-hyperb}
\tilde g ^\H _{dS} = -dT^2 + \sinh^2 \frac{T}{\k}\left(dR^2+\k^2\sinh^2
\frac{R}{\k}\left(d\theta^2+\sin^2\theta d \psi^2\right)\right).
 \eeq
In \cite{LXZ}, two positive energy theorems were proved for certain
$\P$-asymptotically de Sitter and $\H$-asymptotically de Sitter initial data sets.
These initial data sets are asymptotic to time slices of the two kinds of half-de Sitter spacetimes
covered by the planar coordinates and hyperbolic coordinates respectively.
And their mean curvatures satisfy (\ref{trace-K}) for $\P$-asymptotically de Sitter initial data sets
and satisfy (\ref{traceH-K}) for $\H$-asymptotically de Sitter initial data sets.

In the case of zero or negative cosmological constants, the positive energy theorem holds for any
asymptotically flat or asymptotically anti-de Sitter spacelike hypersurfaces whose mean curvatures
have no restriction as (\ref{trace-K}) or (\ref{traceH-K}). This motivates us to study further the special
feature of the conditions (\ref{trace-K}) and (\ref{traceH-K}) in the case of positive cosmological constant.
In this short paper, we shall construct spacelike hypersurfaces with negative total energy in the two kinds of
half-de Sitter spacetimes. The constructed spacelike hypersurfaces are either $\P$-asymptotically de Sitter or
$\H$-asymptotically de Sitter. However, their mean curvatures violate (\ref{trace-K})
in the $\P$-asymptotically de Sitter case and violate (\ref{traceH-K}) in the $\H$-asymptotically de Sitter case.
As the two kinds of half-de Sitter spacetimes satisfy the vacuum Einstein field equations with positive cosmological
constant, the dominant energy condition holds automatically. So these examples provide counterexamples of the positive
energy theorem for general asymptotically de Sitter spacelike hypersurfaces. Thus it indicates that the positive
energy theorem is no longer the general feature in theory of gravity when the cosmological constant is positive.

We would like to point out that de Sitter spacetime can be also separated into two parts
along the hypersurface $X^3=0$. The time slices in this half-de Sitter spacetime are hemispheres.
The counterexample to Min-Oo's conjecture related to the rigidity of hemispheres, constructed by
Brendle, Marques and Neves, can be understood as the another failure of the positive energy theorem
for spacetimes with positive cosmological constant \cite{BMN}. We refer to the survey
paper \cite{B} for the recent works on Min-Oo's conjecture.

\mysection{Positive energy theorems} \ls

In this section, we review two positive energy theorems proved in \cite{LXZ}.
Let $(N^{1,3},\tilde{g})$ be a spacetime satisfying the Einstein equations
with a positive cosmological constant $\Lambda=\frac{3}{\k ^2}$.
 Let $(M, g, K)$ be a spacelike hypersurface with induced Riemannian metric $g$
and second fundamental form $K$. It is $\P$-asymptotically de Sitter of
order $\tau >\frac{1}{2}$ if there is a compact set $M _c$ such that
$M -M _c$ is the disjoint union of a finite number of subsets $M
_1$, $\cdots$, $M _k$ - called the ``ends" of $M$ - each
diffeomorphic to $\R ^3 -B_r$ where $B _r$ is the closed ball of
radius $r$ with center at the coordinate origin. And $g$ and $h=K -\sqrt{\frac{\Lambda}{3}} g$
satisfy
 \beQ \label{g}
 g= \P ^2 \bar g, \quad h= \P \bar h, \quad
\g -  \breve{g} \in C ^{2, \alpha} _{-\tau}(M),\quad
\h \in  C ^{1, \alpha} _{-\tau -1}(M)
 \eeQ
for certain $\tau >\frac{1}{2}$, $0 <\alpha < 1$, where $\P$ is certain positive constant,
$\breve{g}$ is the standard metric of $\R ^3$, $C ^{k,\alpha} _{-\tau}$ are
weighted H\"{o}lder spaces defined in \cite{Ba}. Moreover, the scalar curvature $\cR _g \in L^1(M)$,
$T_{0i} \in L^1(M)$. Let the metric $\breve{g} _{\P } =\P ^2 \breve{g}$.
Let $\{x^i\}$ be natural coordinates of $\R ^3$, $\g _{ij}=\g\left(\pa _i, \pa
_j\right)$, $\h _{ij}=\h\left(\pa _i, \pa _j\right)$.
The total energy $E_l$ and the total linear momentum $P _{lk}$ of $M_l$
for $\P $-asymptotically de Sitter initial data set $(M, g, K)$ are
 \beQ \label{definition}
 \begin{aligned}
 E _l &=\frac{\P}{16\pi}\lim _{r \rightarrow \infty} \int
_{S_{r,l}}\left(\partial _j \g _{ij}-\partial _i \g _{jj}\right)*dx ^i,
\label{bEl}\\
 P _{lk} &= \frac{\P ^2}{8\pi}\lim _{r \rightarrow \infty}
\int _{S_{r,l}}\left(\h_{ki}-\g _{ki} tr_{\g} \left(\h \right)\right)*dx ^i.\label{bPlk}
 \end{aligned}
 \eeQ

An initial data set $(M, g, K)$ is $\H $-asymptotically
de Sitter of order $\tau >\frac{3}{2}$ if there is a compact set $M _c$ such that
$M -M _c$ is the disjoint union of a finite number of subsets $M
_1$, $\cdots$, $M _k$ - called the ``ends" of $M$ - each
diffeomorphic to $\R ^3 -B_r$ where $B _r$ is the closed ball of
radius $r$ with center at the coordinate origin. And $g$ and $h=K-\frac{\coth\frac{T}{\k}}{\k}g$ satisfy
 \beQ
 \begin{aligned}
g=\H^2 \bar g,\quad h=\H \bar h,\quad a_{ij}=O\left(e^{-\frac{\tau}{\k}r}\right), \quad \hconn ^\H _k
a_{ij}=O\left(e^{-\frac{\tau}{\k}r}\right), \\
\quad \hconn ^\H _l \hconn ^\H _k a_{ij}=O\left(e^{-\frac{\tau}{\k}r}\right),\quad
\bar h _{ij}=O\left(e^{-\frac{\tau}{\k}r}\right), \quad \hconn ^\H _k \bar h
_{ij}=O\left(e^{-\frac{\tau}{\k}r}\right)
 \end{aligned}
 \eeQ
on each end for certain constant $T\neq 0$, where $\H=\sinh\frac{T}{\k}$,
$a_{ij}=\bar g\left(\breve{e} _i,\breve{e} _j\right)-\breve{g}_\H\left(\breve{e}_i,\breve{e}_j\right)$,
$\bar h_{ij}=\bar h \left(\breve{e} _i,\breve{e} _j \right)$. And $\breve{g} _{\H}$ is the
hyperbolic metric
 \beQ
\breve{g} _{\H} =dR^2+\k^2\sinh^2\frac{R}{\k}\left(d\theta^2+\sin^2\theta
d \psi^2\right). \label{gH}
 \eeQ
$\breve{e} _1=\partial _R$, $\breve{e}_2=\frac{\partial _\theta}{\k\sinh\frac{R}{\k}}$,
$\breve{e}_3=\frac{\partial_\psi}{\k\sinh\frac{R}{\k}\sin\theta}$ is the frame, $\{\breve{e} ^i\}$ is the coframe.
$\hconn ^\H$ is the Levi-Civita connection of $\breve{g}_\H$. Let $\bar {\cR}$, $\bar \nabla$, $\rho_z$ be
the scalar curvature, the Levi-Civita connection of $\bar g$ and the distance function with respect to $z \in M$
respectively. We further require
$\left(\bar {\cR} +\frac{6}{\k^2}\right)e^{\frac{\rho_z}{\lambda}} \in L^1(M)$,
$\left(\bar \nabla^j \bar h_{ij}-\bar \nabla_i tr_{\bar g}\left(\bar h\right)\right)e^{\frac{\rho_z}{\lambda}} \in
L^1(M)$. Denote
 \beQ
\mathcal{E} _l =\hconn ^{\H,j} \bar g_{1j}- \hconn ^{\H} _1 tr
_{\breve{ g} _\H}\left(\bar g \right) +\frac{1}{\k}\left(a_{22}+a_{33}\right)+2\left(\bar h
_{22}+\bar h _{33}\right).
 \eeQ
The total energy-momentum of the end $M _l$ are
 \beQ
E^{\H} _{l \nu}=\frac{\H ^2}{16\pi}\lim_{R \rightarrow
\infty}\int_{S_{R,l}}\mathcal{E} _l n^\nu e^{\frac{R}{\k}} \breve{e}^2\wedge \breve{e}^3
 \eeQ
where $0\leq \nu \leq 3$, $\left(n^\nu \right)=\left(1, \sin\theta \cos\psi, \sin\theta \sin\psi, \cos\theta\right)$,
$S_{R,l}$ is the coordinate sphere of radius $R$ in $M_l$.

The following two positive energy theorems are proved in \cite{LXZ}.

 \begin{thm}\label{dspmtsym}
Let $(M, g, K)$ be an $\P $-asymptotically de Sitter initial data set
of order $1 \geq \tau >\frac{1}{2}$ which has possibly a finite
number of apparent horizons in spacetime $(N^{1,3},\tilde{g})$ with
positive cosmological constant $\Lambda
>0$. Suppose $N^{1,3}$ satisfies the dominant energy condition.
If the trace of the second fundamental form of $M$ satisfies
 \beq
tr _g (K) \leq \sqrt{3\Lambda }, \label{trace-K}
 \eeq
then for each end $M _l$,
 \beQ
E _l \geq \left|P _{l}\right|_{\breve{g} _{\P}}. \label{dS-pmt}
 \eeQ
If $E_l=0$ for some end $M _{l _0}$, then
 \beQ
\left(M, g, K\right) \equiv \left(\R ^3, \P ^2 \breve{g},
\sqrt{\frac{\Lambda}{3}} \P ^2 \breve{g}\right).\label{equality2}
 \eeQ
Moreover, the mean curvature achieves the equality in
(\ref{trace-K}) and the spacetime $(N^{1,3}, \tilde g)$ is de Sitter
along $M$. In particular, $(N^{1,3}, \tilde g)$ is globally de
Sitter in the planar coordinates if it is globally hyperbolic.
 \end{thm}

\begin{thm}\label{dspmtsymhy}
Let $(M, g, K)$ be an $\H $-asymptotically de Sitter initial data
set of order $\tau >\frac{3}{2}$ which has possibly a finite number
of apparent horizons in spacetime $(N^{1,3},\tilde{g})$ with
positive cosmological constant $\Lambda
>0$. Suppose $N^{1,3}$ satisfies the dominant energy condition. If
the trace of the second fundamental form of $M$ satisfies
 \beq
tr _g (K) \sinh\frac{T}{\lambda} \leq \sqrt{3\Lambda } \cosh\frac{T}{\lambda}, \label{traceH-K}
 \eeq
then for each end $M _l$,
 \beQ
E^\H _{l0}\geq \sqrt{\left(E^\H _{l1}\right)^2 +\left(E^\H _{l2} \right)^2 +\left(E^\H _{l3}\right) ^2}.\label{ds-pmt3}
 \eeQ
If $E^\H _{l0}=0$ for some end $M _{l _0}$, then
 \beQ
\left(M, g, K\right) \equiv \left(\hbox{\bb H} ^3, \sinh^2 \frac{T}{\lambda} \breve{g}_{\H},
\sqrt{\frac{\Lambda}{3}} \sinh\frac{T}{\lambda}\cosh\frac{T}{\lambda}\breve{g}_{\H} \right).\label{equality3}
 \eeQ
Moreover, the mean curvature achieves the equality in
(\ref{traceH-K}) and the spacetime $(N^{1,3}, \tilde g)$ is de
Sitter along $M$. In particular, $(N^{1,3}, \tilde g)$ is globally
de Sitter in the hyperbolic coordinates if it is globally
hyperbolic.
\end{thm}

In general, the conformal factors $\P$ and $\H$ are functions on 3-manifolds. The above positive energy theorems
hold only when $\P$ and $\H$ are constants on ends. Let $e_0$ be timelike unit normal to $M$. It is well-known that
\beQ
K =\frac{1}{2}\left(\Lie _{e_0} \tilde g \right)^{proj}
\eeQ
where $\Lie $ is the Lie derivative and ``$proj$" means the projection to the tangent bundle of $M$. For $t$-slice
with $e_0=\partial _t$ in the metric (\ref{dS-planar}),
\beQ
tr_g(K) =\sqrt{3\Lambda },
\eeQ
and for $T$-slice ($T\neq 0$) with $e_0=\partial _T$ in the metric (\ref{dS-hyperb}),
\beQ
tr_g(K)=\sqrt{3\Lambda } \coth\frac{T}{\lambda}.
\eeQ
Thus the mean curvatures conditions (\ref{trace-K}) and (\ref{traceH-K}) indicate that, in certain sense of average,
the 3-space evolution in spacetime $N$ should not be too rapid to exceed the standard de Sitter spacetime in order to
keep positivity of the total energy.

\mysection{Negative total energy} \ls

In this section, we construct certain spacelike hypersurfaces with negative total energy. The mean
curvatures of these hypersurfaces violate the condition (\ref{trace-K}) or (\ref{traceH-K}).
These hypersurfaces are constructed by slightly perturbing the $t$ or $T$-slices in half-de Sitter
spacetime.

First of all, we provide the formula of the second fundamental form of a graph.
Let $(M^n, \hat{g})$ be an $n$-dimensional Riemannian manifold, $\cconn $ be its Levi-Civita connection,
$I$ be an interval of $\hbox{\bb R}$. We equip $I\times M$ with the wrapped product
 \beQ
\tilde{g}=-dt^2+\rho^2\hat{g},
 \eeQ
where $\rho=\rho(t)$ is a positive smooth function. Suppose that $\Sigma $ is a smooth graph which is
given by
 \beQ
\begin{aligned}
F:& M &&\longrightarrow \quad  I\times M\\
  & x &&\longmapsto \quad (f(x),x).
\end{aligned}
\label{GraphP}
 \eeQ
Choosing a frame $\{\hat{e}_i\}$ on $M$, the tangent space of the graph $\Sigma$
is spanned by
 \beQ
F_*\hat{e}_i=\cconn _i f \partial _t+\hat{e}_i, \qquad i=1,\ldots,n.
 \eeQ
As $\partial _t + \rho ^{-2}\cconn f$ is orthogonal to $\Sigma$, the graph is spacelike if and only if
 \beq
\rho ^2 > |\cconn f|^2 _{\hat{g}}. \label{condofspacelike}
 \eeq
Then,
$\nu =\frac{\partial _t + \rho ^{-2}\cconn f}{\sqrt{1-\rho ^{-2}|\cconn f|^2 _{\hat{g}}}}$
is a timelike unit normal vector field of the graph. Let $g$ be the induced metric on $\Sigma$.
The components of the second fundamental form with respect to the normal
vector $\nu$ are
 \beQ
 \begin{aligned}
K_{ij}=&K(F_*\hat{e}_i,F_*\hat{e}_j)
       =\mu \left(\cconn ^2 _{i,j} f - 2 \rho ' \rho ^{-1}\cconn _i f \cconn _j f + \rho ' \rho \hat{g}_{ij}\right) \label{expofK},\\
 tr _g (K)=&\mu \,\rho ^{-2}\left(\hat{g}^{ij}+\mu ^{2}\rho ^{-2} \cconn^i f \cconn ^j f \right)\cconn ^2 _{i,j} f
 -\mu ^3 \rho ' \rho ^{-1}+(n+1) \mu  \rho ' \rho ^{-1},\label{expofH}
  \end{aligned}
  \eeQ
where
$\mu =\left(1-\rho ^{-2}|\cconn f|^2 _{\hat{g}}\right)^{-\frac{1}{2}} $.

{\em (i) The $\P$-asymptotically de Sitter case}: Consider the graph where
 \beQ
f(x)=t_0+\varepsilon \left(1+r^2 \right)^{-\frac{1}{2}}
 \eeQ
for certain constant $t_0$ in the half-de Sitter spacetime equipped with the metric (\ref{dS-planar}) in the planar
coordinates. As
 \beQ
\rho ^{-2} |\hconn f|^2 _{\breve{g}} = \varepsilon ^2\left(1+r^2\right)^{-3}r^2 e^{-\frac{2}{\k}\left(t_0+\varepsilon (1+r^2)^{-\frac{1}{2}}\right)},
 \eeQ
(\ref{condofspacelike}) holds and the graph is an embedding spacelike hypersurface for $\varepsilon$ is sufficiently small.

Note that
\beQ
 \begin{aligned}
  \partial _i f =& - \varepsilon \left(1+r^2 \right)^{-\frac{3}{2}}x_i,\\
  \partial^2_{i,j}f =& 3\varepsilon \left(1+r^2 \right)^{-\frac{5}{2}}x_ix_j - \varepsilon \left(1+r^2 \right)^{-\frac{3}{2}}\delta_{ij},
 \end{aligned}
 \eeQ
thus the induced metric and the second fundamental forms of the graph are
 \beQ
 \begin{aligned}
g_{ij}=& \,e^{\frac{2f}{\k}}\delta _{ij} - \varepsilon ^2 \left(1+r^2 \right)^{-3} x_i x_j,\\
K_{ij}=& \,\mu \left(\frac{1}{\k}e^{\frac{2f}{\k}}\delta _{ij} - \varepsilon \left(1+r^2\right)^{-\frac{3}{2}}\delta _{ij}
        -\frac{2}{\k}\varepsilon ^2 \left(1+r^2 \right)^{-3} x_i x_j + 3\varepsilon \left(1+r^2 \right)^{-\frac{5}{2}} x_i x_j\right),
 \end{aligned}
 \eeQ
where $\mu = \left(1-\varepsilon ^2 r^2 (1+r^2)^{-3}e^{-\frac{2f}{\k}}\right)^{-\frac{1}{2}} = 1 + \frac{1}{2}\varepsilon ^2 r^2 (1+r^2)^{-3}e^{-\frac{2f}{\k}}+ O(r^{-8}).$
Set $\P=e^{\frac{t_0}{\k}}$, $\bar{g} = \P^{-2}g$, $a = \bar{g}- \breve {g}$ and $\bar{h} = \P^{-1}\left(K - \frac{1}{\k}g\right)$. We have,
 \beQ
 \begin{aligned}
 a_{ij}=& \,\left(e^{\frac{2\left(f-t_0 \right)}{\k}}-1\right)\delta _{ij}-\P^{-2}\varepsilon^2 \left(1+r^2 \right)^{-3}x_i x_j
       = \,\left(e^{\frac{2\left(f-t_0\right)}{\k}}-1 \right)\delta _{ij}+O\left(r^{-4}\right),\\
\partial _k a_{ij}=& \,O\left(r^{-2}\right),\qquad  \partial_k \partial_l a_{ij}=O\left(r^{-3}\right),\\
\bar{h}_{ij}=& \P^{-1} \left(\left( - \frac{1}{\k}e^{\frac{2f}{\k}}\delta _{ij} + O\left(r^{-3}\right) \right) -
                \frac{1}{\k}\left(e^{\frac{2f}{\k}}\delta _{ij} + O\left(r^{-4}\right)\right)\right)= O\left(r^{-3}\right).
\end{aligned}
\eeQ
Using the formula on the difference of the scalar curvatures of two metrics (c.f. \cite{BMN}, p187),
 \beQ \label{DifferenceOfScalarCurvature}
 \begin{aligned}
 \cR _{\bar{g}} =& \cR _{\breve {g}}+\sum _{s=1}^{\infty}(-1)^s \left\langle Ric_{\breve {g}},a^s \right\rangle _{\breve {g}}
                               +\frac{1}{4}\bar{g}^{ik}\bar{g}^{jl}\bar{g}^{pq}
                              \Big(-2\hconn _i a_{lp}\hconn _j a_{kq}-4\hconn _i a_{kp}\hconn _j a_{lq}\\
                              &+4\hconn _i a_{kl}\hconn _j a_{pq}+3\hconn _i a_{lq}\hconn _k a_{jp}-\hconn _i a_{jl}\hconn _k a_{pq}\Big)
                               -\bar{g}^{ik}\bar{g}^{jl}\left(\hconn^2 _{i,k} a_{jl}-\hconn^2 _{i,l} a_{jk}\right),\\
 \end{aligned}
 \eeQ
we obtain
 \beQ
 \begin{aligned}
 \cR _{\bar{g}}=&- \bar{g}^{ik}\bar{g}^{jl}\left(\hconn^2 _{i,k} a_{jl}-\hconn^2 _{i,l} a_{jk}\right) + O\left(r^{-4}\right)\\
            =&-\Delta _{\breve {g}} a_{jj} + \hconn^2 _{i,j} a_{ij} + O\left(r^{-4}\right)\\
            =&-2\Delta _{\breve {g}}e^{\frac{2(f-t_0)}{\k}}+ O\left(r^{-4}\right)\\
            =&-\frac{4}{\k}e^{\frac{2(f-t_0)}{\k}}\Delta _{\breve {g}}f+ O\left(r^{-4}\right)=O\left(r^{-4}\right).
 \end{aligned}
 \eeQ
This implies that $\cR_{g}$ is $L^1$. Thus, $(\R^3, g, K)$ is $\P$-asymptotically de Sitter with the constant $\P=e^{\frac{t_0}{\k}}$. Note that
\beQ \label{MeanCurvOfGraphP}
\begin{aligned}
tr _g (K)=&\frac{4}{\k}\mu - \frac{1}{\k}\mu ^3+
              \mu e^{-\frac{2f}{\k}}
             \left(\mu ^2 e^{-\frac{2f}{\k}}\varepsilon ^2 \left(1+r^2\right)^{-3} x_i x_j + \delta _{ij}\right)\\
           & \cdot \left(-\varepsilon \left(1+r^2\right)^{-\frac{3}{2}}\delta _{ij} +  3\varepsilon \left(1+r^2\right)^{-\frac{5}{2}} x_i x_j\right)\\
         =&\frac{3}{\k} + \frac{1}{2\k}\varepsilon ^2 r^2 \left(1+r^2\right)^{-3}e^{-\frac{2f}{\k}} + O\left(r^{-5}\right).
\end{aligned}
\eeQ
This shows that if $ \varepsilon \neq 0$, then $tr _g (K) > \frac{3}{\k}$ for large $r$, which violates the condition (\ref{dspmtsym}).
Now the total energy is
\beQ \label{EnergyOfGraphP}
 \begin{aligned}
    E &=\frac{\P}{16\pi}\lim _{r \rightarrow \infty} \int
         _{S_r}\left(\partial _j \g _{ij}-\partial _i \g _{jj}\right)*dx ^i\\
      &=\frac{\P}{16\pi}\lim _{r \rightarrow \infty} \int _{S_r}
           \left(\frac{4\varepsilon }{\k}r\left(1+r^2\right)^{-\frac{3}{2}} e^{\frac{2(f-t_0)}{\k}}- 3\P ^{-2}
           \varepsilon ^2 r\left(1+r^2\right)^{-3} \right)d\sigma _r\\
      &=\frac{\varepsilon}{\k}e^{\frac{t_0}{\k}}.
 \end{aligned}
\eeQ
So $E<0$ if $\varepsilon <0$. Furthermore, the total linear momentum
\beQ \label{totallinearmomentumOfGraphP}
 \begin{aligned}
     P _k = \frac{\P ^2}{8\pi}\lim _{r \rightarrow \infty} \int _{S_{r}}\left(\h_{ki}-\g _{ki} tr_{\g} \left(\h \right)\right)*dx ^i
          = \frac{\P ^2}{8\pi}\lim _{r \rightarrow \infty} \int _{S_{r}}\left(O\left(r^{-3}\right)\right)*dx ^i = 0.
 \end{aligned}
\eeQ

{\em (ii) The $\H$-asymptotically de Sitter case}: Consider the graph where
 \beQ
f(x)=T_0+\varepsilon e ^{\frac{-3R}{\k}}
 \eeQ
for certain constant $T_0 \neq 0$ in the half-de Sitter spacetime equipped with the metric (\ref{dS-hyperb}) in the hyperbolic
coordinates. (Note that the hyperbolic coordinates are only valid for $T\neq0$, $T_0 =0$ can not be chosen here.) As
 \beQ
\rho ^{-2} |\hconn^{\H} f|^2 _{\breve{g_\H}} = \frac{9\varepsilon^2}{\k ^2} e ^{\frac{-6R}{\k}}\sinh ^{-2}{\frac{f}{\k}},
 \eeQ
(\ref{condofspacelike}) holds and the graph is an embedding spacelike hypersurface for $\varepsilon$ is sufficiently small.

Recall that the standard hyperbolic metric $\breve g _\H$ with the $\breve g_\H$-orthonormal frame
$\breve{e}_1 $, $\breve{e}_2$, $\breve{e}_3$. By Cartan's structure equations
$d\breve{e}^i=-\omega^i_j\wedge \breve{e}^j$ and $\Gamma^j_{ki}=\omega^i_j(\breve{e}_k)$, we obtain
\beQ
  \omega^2_1 = -\omega^1_2 = \cosh{\frac{R}{\lambda}}d\theta,\quad
  \omega^3_1 = -\omega^1_3 = \cosh{\frac{R}{\lambda}}\sin\theta d\psi,\quad
  \omega^3_2 = -\omega^2_3 = \cos\theta d\psi,\\
  \Gamma^2_{21} = - \Gamma^1_{22} = \frac{1}{\lambda} \coth\frac{R}{\lambda},\quad
  \Gamma^3_{31} = - \Gamma^1_{33} = \frac{1}{\lambda} \coth\frac{R}{\lambda},\quad
  \Gamma^3_{32} = - \Gamma^2_{33} = \frac{\cot\theta}{\lambda \sinh\frac{R}{\lambda}},
\eeQ
and other $\omega^i_j$, $\Gamma^k_{ij}$ vanish. Clearly, the gradient of $f$ is
\beQ
  \hconn^\H f = &\,\left(-\frac{3\varepsilon }{\k}e ^{\frac{-3R}{\k}},0, 0, 0 \right).
\eeQ
Since
\beQ
\begin{aligned}
  \hconn^\H _{1,1} f = & \breve{e}_1\breve{e}_1 f - \breve{e}_i (f)\Gamma^i_{11}
                     =   \partial _R \partial _R f
                     =   \frac{9\varepsilon }{\k ^2}e ^{\frac{-3R}{\k}}, \\
  \hconn^\H _{2,2} f = & \breve{e}_2\breve{e}_2 f - \breve{e}_i (f)\Gamma^i_{22}
                     =   - \breve{e}_1 (f)\Gamma^1_{22}
                     =   -\frac{3\varepsilon }{\k ^2}e ^{\frac{-3R}{\k}}\coth{\frac{R}{\k}}, \\
  \hconn^\H _{3,3} f = & \breve{e}_3\breve{e}_3 f - \breve{e}_i (f)\Gamma^i_{33}
                     =   - \breve{e}_1 (f)\Gamma^1_{33}
                     =   -\frac{3\varepsilon }{\k ^2}e ^{\frac{-3R}{\k}}\coth{\frac{R}{\k}}, \\
  \hconn^\H _{1,2} f = & \breve{e}_1\breve{e}_2 f - \breve{e}_i (f)\Gamma^i_{12}
                     =   0, \\
  \hconn^\H _{1,3} f = & \breve{e}_1\breve{e}_3 f - \breve{e}_i (f)\Gamma^i_{13}
                     =   0, \\
  \hconn^\H _{2,3} f = & \breve{e}_2\breve{e}_3 f - \breve{e}_i (f)\Gamma^i_{23}
                     =   0,\\
\end{aligned}
\eeQ
we can find the Hessian of $f$
\beQ
  \left(\hconn^\H _{i,j} f\right) = \frac{3\varepsilon }{\k ^2}e ^{\frac{-3R}{\k}}
  \mbox{diag} \left(3,-\coth{\frac{R}{\k}},-\coth{\frac{R}{\k}}\right).
\eeQ
Then the induced metric and the second fundamental forms of the graph are
 \beQ
 \begin{aligned}
g =&\left( \sinh^2{\frac{f}{\k}} - \frac{9\varepsilon^2}{\k^2}e ^{\frac{-6R}{\k}} \right)dR^2 + \k^2\sinh^2{\frac{f}{\k}}\sinh^2\frac{R}{\k}\left(d\theta^2+\sin^2\theta \psi^2\right),\\
K_{11}= &\mu \left( \frac{1}{\k}\cosh{\frac{f}{\k}}\sinh{\frac{f}{\k}} +
        \frac{9\varepsilon }{\k ^2}e ^{\frac{-3R}{\k}}- \frac{18\varepsilon ^2}{\k ^3}\coth{\frac{f}{\k}}e ^{\frac{-6R}{\k}}
         \right) ,\\
      = &\frac{1}{\k}\sinh{\frac{T_0}{\k}}\cosh{\frac{T_0}{\k}} + \frac{\varepsilon}{\k^2}\left(9+\sinh^2{\frac{T_0}{\k}}
          +\cosh^2{\frac{T_0}{\k}}\right)e ^{\frac{-3R}{\k}} + O\left(e ^{\frac{-6R}{\k}}\right),\\
K_{22}= &K_{33}
      = \mu\left( \frac{1}{\k}\cosh{\frac{f}{\k}}\sinh{\frac{f}{\k}}
               -\frac{3\varepsilon }{\k ^2}e ^{\frac{-3R}{\k}} \coth{\frac{R}{\k}}\right) \\
      =&\frac{1}{\k}\sinh{\frac{T_0}{\k}}\cosh{\frac{T_0}{\k}} + \frac{\varepsilon}{\k^2}\left(-3+\sinh^2{\frac{T_0}{\k}}
          +\cosh^2{\frac{T_0}{\k}}\right)e ^{\frac{-3R}{\k}} + O\left(e ^{\frac{-5R}{\k}}\right),
 \end{aligned}
 \eeQ
and other $K_{ij}=0$, where $\mu = \left( 1 - \frac{9\varepsilon^2}{\k^2}e ^{\frac{-6R}{\k}}\sinh^{-2}{\frac{f}{\k}} \right) ^{-\frac{1}{2}} =
1 + O\left(e^{\frac{-6R}{\k}}\right).$ Set $\H=\sinh {\frac{T_0}{\k}}$, $\bar{g} = \H^{-2}g$, $a=\bar{g}-\breve{g}_{\H}$ and
$\bar{h}=\H^{-1} \left(K-\frac{\coth{\frac{T_0}{\k}}}{\k}g\right) $, we obtain
 \beQ
 \begin{aligned}
\left( a_{ij} \right)=& \,\frac{2\varepsilon }{\k} \coth {\frac{T_0}{\k}}e^{\frac{-3R}{\k}}
                          \mbox{diag} \left(1+ O(e^{\frac{-3R}{\k}}), 1+ O(e^{\frac{-3R}{\k}}),1+ O(e^{\frac{-3R}{\k}})\right),\\
 \left(\bar{h}_{ij} \right) =& \,\frac{4\varepsilon }{\k^2} \sinh^{-1} {\frac{T_0}{\k}}e^{\frac{-3R}{\k}}
 \mbox{diag} \left(2+ O(e^{\frac{-3R}{\k}}), -1+ O(e^{\frac{-2R}{\k}}),-1+ O(e^{\frac{-2R}{\k}}) \right),\\
 \hconn^\H _{k} a _{ij}= & \,O\left(e^{\frac{-3R}{\k}}\right), \quad \hconn^{\H} _{k,l} a _{ij}=  O\left(e^{\frac{-3R}{\k}}\right), \quad
 \hconn^\H _{k} \bar h _{ij}= O\left(e^{\frac{-3R}{\k}}\right).
 \end{aligned}
 \eeQ
Now we compute the scalar curvature. Note that
\beQ
\begin{aligned}
  tr_{\breve{g}_{\H}}a =& a_{11} + a_{22} + a_{33}
                       = \frac{6\varepsilon}{\lambda}\coth{\frac{T_0}{\lambda}}e^{-\frac{3R}{\lambda}} +  O(e^{\frac{-6R}{\lambda}}),\\
  \hconn^{\H} _{i,i} a _{jj} =& \frac{6\varepsilon}{\lambda}\coth{\frac{T_0}{\lambda}}\Delta_{\breve{g}_{\H}}e^{-\frac{3R}{\lambda}} +
                                 O\left(e^{\frac{-6R}{\lambda}}\right),\\
  \hconn^{\H} _{i,j} a _{ij} =& \hconn^{\H} _{i,j} \left(  \frac{2\varepsilon}{\lambda}\coth{\frac{T_0}{\lambda}}e^{-\frac{3R}{\lambda}}
                                 \breve{g}_{\H ij} + O(e^{\frac{-6R}{\lambda}})\right)
                             = \frac{2\varepsilon}{\lambda}\coth{\frac{T_0}{\lambda}}\Delta_{\breve{g}_{\H}}e^{-\frac{3R}{\lambda}}
                             +O\left(e^{\frac{-6R}{\lambda}}\right),\\
  \Delta_{\breve{g}_{\H}}e^{\frac{-3R}{\lambda}} = &\frac{3}{\lambda^2}e^{-\frac{3R}{\lambda}} + O\left(e^{\frac{-5R}{\lambda}}\right),
\end{aligned}
\eeQ
We have
\beQ
\begin{aligned}
  \cR _{\bar{g}}=&\cR _{\breve{g}_{\H}} - \left\langle Ric_{\breve{g}_{\H}}, a \right\rangle _{\breve{g}_{\H}}
               - \hconn^{\H} _{i,i} a _{jj} + \hconn^{\H} _{i,j} a _{ij}+O\left(e^{\frac{-6R}{\k}}\right)\\
            =&-\frac{6}{\k^2}+\frac{12\varepsilon}{\k ^3}\coth{\frac{T_0}{\k}}e^{-\frac{3R}{\k}}
              -\frac{18 \varepsilon}{\k ^3}\coth{\frac{T_0}{\k}}e^{-\frac{3R}{\k}}
             +\frac{6 \varepsilon}{\k ^3}\coth{\frac{T_0}{\k}}e^{-\frac{3R}{\k}}
              +O\left(e^{\frac{-5R}{\k}}\right)\\
            =&-\frac{6}{\k^2}+O\left(e^{\frac{-5R}{\k}}\right).
\end{aligned}
\eeQ
So $\left(\cR _{\bar{g}} +\frac{6}{\k^2}\right)e^{\frac{\rho_z}{\lambda}}$ is $L^1$ with respect to the metric $\bar{g}$.
Next, we shall show
that $\left(\bar \nabla^j \bar h_{ij}-\bar \nabla_i tr_{\bar g}\left(\bar h\right)\right)e^{\frac{R}{\lambda}}$ is $L^1\left(\mathbb{R}^3,\breve{g}_{\H}\right)$. As
\beQ
\begin{aligned}
  \hconn^{\H}_i tr_{\breve{g}_{\H}}(\bar h) =& O\left(e^{\frac{-5R}{\lambda}}\right), \\
  \sum_j \hconn^{\H}_{j} \bar h_{1j} =& \breve{e}_j \left(\bar{h}_{1j}\right)-\bar{h}_{kj}\Gamma^k_{j1}-\bar{h}_{1k}\Gamma^k_{jj}\\
                                   =& \breve{e}_1 \left(\bar{h}_{11}\right) - \bar{h}_{22}\Gamma^2_{21} - \bar{h}_{33}\Gamma^3_{31}-\bar{h}_{11}\left(\Gamma^1_{22}+\Gamma^1_{33}\right)\\
                                   =& \partial _R \left(\frac{8\varepsilon }{\k^2} \sinh^{-1} {\frac{T_0}{\k}}e^{\frac{-3R}{\k}}\right)
                                      +\frac{4\varepsilon }{\k^2} \sinh^{-1} {\frac{T_0}{\k}}e^{\frac{-3R}{\k}}\left(\frac{2}{\lambda} \coth\frac{R}{\lambda} \right)\\
                                    &-\frac{8\varepsilon }{\k^2} \sinh^{-1} {\frac{T_0}{\k}}e^{\frac{-3R}{\k}}\left(-\frac{2}{\lambda} \coth\frac{R}{\lambda}\right)+ O\left(e^{\frac{-6R}{\lambda}}\right)
                                   =O\left(e^{\frac{-5R}{\lambda}}\right),\\
  \sum_j \hconn^{\H}_{j} \bar h_{2j} =& \breve{e}_j \left(\bar{h}_{2j}\right)-\bar{h}_{kj}\Gamma^k_{j2}-\bar{h}_{2k}\Gamma^k_{jj}\\
                                     =& -\bar{h}_{jj}\Gamma^j_{j2} - \bar{h}_{22}\Gamma^2_{jj}+O\left(e^{\frac{-5R}{\lambda}}\right)
                                     = O\left(e^{\frac{-5R}{\lambda}}\right),\\
  \sum_j \hconn^{\H}_{j} \bar h_{3j} =& \breve{e}_j \left(\bar{h}_{3j}\right)-\bar{h}_{kj}\Gamma^k_{j3}-\bar{h}_{3k}\Gamma^k_{jj}\\
                                     =& -\bar{h}_{jj}\Gamma^j_{j3} - \bar{h}_{33}\Gamma^3_{jj}+O\left(e^{\frac{-5R}{\lambda}}\right)
                                     = O\left(e^{\frac{-5R}{\lambda}}\right),
\end{aligned}
\eeQ
we obtain that $\left(\breve \nabla^{\H j} \bar h_{ij}-\breve \nabla^{\H}_i tr_{\breve g_{\H}}\left(\bar h\right)\right)e^{\frac{R}{\lambda}}$
is $L^1\left(\mathbb{R}^3,\breve{g}_{\H}\right)$. Denote symmetric $2$-tensor $\Gamma(X,Y)=\bar \nabla_X Y - \breve \nabla^{\H}_X Y$. There is
constant $C_1$ such that
$\left|\Gamma\right|_{\breve{g}_{\H}}\leq C_1\frac{\left|\breve \nabla^{\H}a\right|_{\breve{g}_{\H}}}{1-\left|a\right|_{\breve{g}_{\H}}}=O\left(e^{\frac{-3R}{\lambda}}\right)$
(c.f. \cite{BMN}, p188). Therefore,
\beQ
\left|\bar \nabla \bar h - \breve \nabla^{\H} \bar h\right|_{\breve{g}_{\H}} \leq C_2 \left|\bar h\right|_{\breve{g}_{\H}}\left|\Gamma\right|_{\breve{g}_{\H}}=O\left(e^{\frac{-6R}{\lambda}}\right),
\eeQ
and we conclude that $\left(\bar \nabla^j \bar h_{ij}-\bar \nabla_i tr_{\bar g}\left(\bar h\right)\right)e^{\frac{R}{\lambda}}$ is $L^1\left(\mathbb{R}^3,\breve{g}_{\H}\right)$. Thus $(\R^3, g, K)$ is $\H$-asymptotically de Sitter with
the constant $\H=\sinh {\frac{T_0}{\k}}$. Now
\beQ
\begin{aligned}
tr _g (K)=&\mu \sinh^{-2}{\frac{f}{\k}}
             \left(\delta _{ij} + \mu ^2 \sinh^{-2}{\frac{f}{\k}}\hconn ^\H _i f \hconn ^\H _j f \right)\hconn^{\H } _{i,j} f \\
          & - \frac{1}{\k}\mu^3\coth {\frac{f}{\k}} + \frac{4}{\k}\mu\coth {\frac{f}{\k}}\\
         =&\frac{3}{\k}\coth {\frac{T_0}{\k}} - \frac{12\varepsilon }{\k ^2}\sinh^{-2}{\frac{T_0}{\k}}
           e ^{\frac{-5R}{\k}}+  O\left(e^{\frac{-6R}{\k}}\right).
\end{aligned}
\eeQ
Therefore,
\beQ
tr _g (K)\sinh{\frac{T_0}{\k}} - \frac{3}{\k}\cosh {\frac{T_0}{\k}} =
   -\frac{12\varepsilon }{\k ^2\sinh{\frac{T_0}{\k}}}e ^{\frac{-5R}{\k}}+ O\left(e^{\frac{-6R}{\k}}\right).
\eeQ

Now we calculate the total energy-momentum. As
\beQ
 \begin{aligned}
    \mathcal{E} =&\hconn ^{\H,j} \bar g_{1j}- \hconn ^{\H} _1 tr_{\breve{ g} _\H}(\bar g) +
                     \frac{1}{\k}(a_{22}+a_{33})+2(\bar h_{22}+\bar h _{33}) \\
                   =&-\frac{6 \varepsilon}{\k ^2}\coth{\frac{T_0}{\k}}e^{-\frac{3R}{\k}} +
                     \frac{18 \varepsilon}{\k ^2}\coth{\frac{T_0}{\k}}e^{-\frac{3R}{\k}} \\
                   &+\frac{4 \varepsilon}{\k ^2}\coth{\frac{T_0}{\k}}e^{-\frac{3R}{\k}} -
                      \frac{16 \varepsilon}{\k ^2}\sinh^{-1}{\frac{T_0}{\k}}e^{-\frac{3R}{\k}} +
                     O\left(e^{\frac{-5R}{\k}}\right) \\
                   =&\frac{16 \varepsilon}{\k ^2}\tanh{\frac{T_0}{2\k}}e^{-\frac{3R}{\k}} + O\left(e^{\frac{-5R}{\k}}\right),
 \end{aligned}
\eeQ
we have
\beQ
 \begin{aligned}
E^{\H} _\nu= &\frac{\H ^2}{16\pi}\lim_{R \rightarrow \infty}\int_{S_R}
              \mathcal{E} n^\nu e^{\frac{R}{\k}} \breve{e}^2\wedge \breve{e}^3 \\
          = &\frac{\sinh^2 {\frac{T_0}{\k}}}{16\pi}\lim_{R \rightarrow \infty}\int_{S_R}
              \frac{16 \varepsilon}{\k ^2}\tanh{\frac{T_0}{2\k}}e^{-\frac{3R}{\k}}n^\nu e^{\frac{R}{\k}}\k^2 \sinh^2 {\frac{R}{\k}}\sin{\theta}d\theta \wedge d\psi \\
          = & \frac{\varepsilon}{4\pi}\tanh{\frac{T_0}{2\k}} \sinh^2 {\frac{T_0}{\k}}\int_{S_1}n^\nu \sin{\theta}d\theta \wedge d\psi.\\
\end{aligned}
\eeQ
Since $\left(n^\nu \right)=\left(1, \sin\theta \cos\psi, \sin\theta \sin\psi, \cos\theta\right)$, we obtain
\begin{subnumcases}
{E^{\H} _\nu=}
\varepsilon\tanh{\frac{T_0}{2\k}} \sinh^2 {\frac{T_0}{\k}}, \quad \mbox{if }\; \nu=0, \nonumber\\
0 ,\qquad \qquad \qquad \qquad \mbox{if } \; \nu = 1,2,3. \nonumber
\end{subnumcases}
So $E^{\H} _0<0$ if $\varepsilon T_0 < 0$ and in this case (\ref{dspmtsymhy}) does not hold for large $R$.

We remark that the de Sitter metric (\ref{dS-hyperb}) collapses at $T=0$ which can be viewed as the big bang
of the universe. It is interesting to find, at the big bang, that
 \beQ
\lim _{T _0 \rightarrow 0} E^{\H} _0 =0.
 \eeQ

\bigskip

{\footnotesize {\it Acknowledgement.} The authors would like to thank the referee for some valuable
comments and suggestions. This work is supported partially by
the National Science Foundation of China (grants 10725105, 10731080, 11021091).}

\end{document}